
\magnification=\magstep 1
\baselineskip=16pt
\hsize=15.5truecm
\vsize=23truecm
\font\texto=cmr10 
\font\titulo=cmbx10 scaled \magstep 1

\def\ii{\'\i}

\overfullrule=0pt

\newbox\Ancha
\def\gros#1{{\setbox\Ancha=\hbox{$#1$}%
\kern-.025em\copy\Ancha\kern-\wd\Ancha
\kern.05em\copy\Ancha\kern-\wd\Ancha
\kern-.025em\raise.0433em\box\Ancha}}

\texto

\

\

\centerline{\titulo VARIATIONAL FORMULATION OF}
\centerline{\titulo  LINEAR TIME-DEPENDENT INVARIANTS \footnote{$^\ast$}{\texto
Work supported in part by project UNAM-DGAPA IN103091. }}

\

\centerline{\it O. Casta\~nos, R. L\'opez-Pe\~na and V.I. Man'ko
\footnote{$^\dagger$}{\texto On leave from Lebedev Institute of
Physics, Moscow, Russia.}}
\centerline{\it Instituto de Ciencias Nucleares, UNAM }
\centerline{\it Apdo. Postal 70-543, 04510 M\'exico, D. F., M\'exico}

\

\

\

{\leftskip .5truecm \rightskip .5truecm
\noindent
{\bf Abstract}.  It is shown that linear time-dependent invariants for
arbitrary multi\-dimensional quadratic systems can be obtained from the
Lagrangian and Hamiltonian formulation procedures by considering a variation
of
coordinates and momenta that follows the classical trajectory and defines a
noetherian symmetry transformation. \smallskip}

\vfill
\eject

\noindent
There are several physical systems which in different approximations
can be represented by Hamiltonians which are quadratic forms of
coordinates and momenta variables, {\it i.e.}, the classical small
oscillations problem. Thus the applicability of multidimensional
quadratic Hamiltonians ranges from solid-state physics to particle
physics, as for example in the relativistic oscillator models of
elementary particles [1,2] and the particle creation in
non-stationary metrics [3].  The exact solutions for stationary
multidimensional quadratic systems have been found and discussed in
the framework of canonical transformations [4].

Recently for the classical small oscillations problem by means of
non-noetherian transformations, noetherian quadratic time-independent
integrals of motion have been found~[5]. For the one-mode parametric
classical oscillator Ermakov found an invariant, which is quadratic in
the momentum and position variables~[6]. In the last years on the base
of a geometric approach related to the Noether's theorem the last
invariant has been rederived [7].

In the case of non-stationary quantum systems there are different
procedures for solving time-dependent Hamiltonians and calculating
evolution operators, but the simplest, at least for quadratic systems,
is the method of time-dependent integrals of motion [8].

The theory of integrals of motion which do not depend on time
in Schr\"odinger representation is well known [9].  On the other
hand, for the one-dimensional quantum oscillator a time-dependent
integral of motion quadratic in position and momentum operators has
been found [10], which in form coincides with the Ermakov invariant.
The linear time-dependent invariants have been found and applied to
get the solutions for the quantum parametric oscillator and a charge
moving under a time-dependent magnetic field [11].  In the most
general multidimentional case, the time-dependent invariants of the
non-stationary quadratic quantum systems also have been found and used
to get the solutions of the time-dependent Schr\"{o}dinger equation
in the form of gaussian coherent states and Hermite polynomials of
several variables [8, 12].  These invariants are linear
combinations of coordinates and momenta operators, with
time-dependent coefficients.

We think it is necessary to understand the existence of these
time-dependent constants of motion within the framework of a
variational formulation. Up to now for one-mode parametric oscillator,
the linear invariants of Ref. [11], were obtained in [13], [14] and
[15] through the Noether's theorem procedure.  Also with this
formalism in Refs. [13] and [14] the quadratic time-dependent
invariants of this system were derived.  Nevertheless until now it was
unknown what symmetry corresponds to linear time-dependent integrals
of motion in the frame of Noether's theorem prescription for
multidimentional parametric oscillators.

The aim of this paper is to get the linear multidimensional
time-dependent invariants by means of the Lagrangian and Hamiltonian
formalisms of the Noether's theorem procedure. We give explicitly such
a variation or noetherian symmetry transformation, which has a clear
physical interpretation. Preliminary results, of the Lagrangian
Noether's theorem procedure, for an arbitrary mutidimensional forced
parametric oscillators with a detailed discussion of a two dimensional
quantum system was presented in Ref.  [15].

Next we present the active viewpoint of the Lagrangian and Hamiltonian
formalisms of the Noether's theorem procedure for a given Hamiltonian system.

\vskip 1.0pc

\noindent
{\bf Lagrangian Formulation}

\vskip 1.0pc

Let us consider an arbitrary time-dependent Hamiltonian, which is
given in terms of a general quadratic form, {\it i.e.,}
$$
H = {1\over 2} Q_{a} {\cal B}_{a b} (t) Q_{b} + {\cal
C}_{a} (t) Q_{a} ~~ , \eqno(1)
$$
where hereafter latin indices run from $ 1 $ to $ 2 n $ and sum over
repeated indices is understood. Besides, we defined the coordinates
and momenta like a $2n$-dimensional vector
$$
Q = \pmatrix{ p_1 \cr \vdots \cr p_n \cr q_1 \cr \vdots \cr q_n \cr}
{}~~ . \eqno(2)
$$
The $2n$-dimensional matrix ${\cal B}$ and vector ${\cal C}$, which
define the quadratic form of the Hamiltonian, can be rewritten in
terms of the $n \times n$ submatrices, $A$, $B$, $C$, and $D$, and the
$n-$column vectors $F$, and $G$, respectively
$$
{\cal B} = \pmatrix{ A & B \cr C & D \cr} ~~ , \qquad {\cal C} =
\pmatrix { F \cr G \cr} ~~ . \eqno(3)
$$
The hermiticity of the Hamiltonian implies that the matrix ${\cal B}$
must be symmetric, and this gives rise to the following symmetry
conditions over the four constituents $n \times n$ matrices:
$$
A^t = A, \quad B^t = C, \quad D^t = D ~~ . \eqno(4)
$$
Expanding the Hamiltonian (1) we obtain
$$
H = {1\over 2} (A_{\alpha\beta} p_\alpha p_\beta  + 2B_{\alpha \beta}
p_\alpha  q_\beta + D_{\alpha\beta} q_\alpha q_\beta) + F_{\alpha}
p_\alpha  + G_\beta q_\beta ~~ , \eqno(5)
$$
where the symmetry conditions (4)  were used.  Now and hereafter, greek indices
run from $ 1 $ to $ n $.  From Hamiltonian (5),
it is straightforward to obtain the Hamilton equations of motion
$$
\eqalignno{
\dot p_\alpha  & = - C_{\alpha \beta} p_\beta - D_{\alpha\beta} q_\beta
- G_\alpha ~~ , & (6a)  \cr
\dot q_\alpha & = A_{\alpha\beta} p_\beta + B_{\alpha\beta} q_\beta +
F_\alpha  ~~ . & (6b) \cr}
$$
This coupled set of first order differential equations can be
decoupled in the following sets of second order equations
$$
\eqalignno{\ddot p_\alpha = & - \biggl[ C - \biggl( \dot D - DB
\biggr) D^{-1} \biggr]_{\alpha\beta} \dot p_\beta - \biggl[ \dot C +
DA - \biggl( \dot D + DB\biggr) D^{-1} C \biggr]_{\alpha\beta}
p_\beta \cr
& + \biggl[ \biggl( \dot D + DB\biggr) D^{-1} \biggr]_{\alpha\beta}
G_\beta + D_{\alpha\beta} F_\beta + \dot G_\alpha ~~ , & (7a) \cr
\ddot q_\alpha = & \biggl[ B + ( \dot A - AC \biggr) A^{-1}
\biggr]_{\alpha\beta} \dot q_\beta + \biggr[ \dot B - AD - \biggl (
\dot A - AC \biggr) A^{-1} B \biggr]_{\alpha\beta} q_\beta \cr
& - \biggl( \biggl[ \dot A - AC \biggr] A^{-1} \biggr)_{\alpha\beta}
F_\beta - A_{\alpha\beta} G_\beta + \dot F_\alpha ~~ . & (7b) \cr}
$$

Next, we are going to find the Lagrangian of the system and apply
Noether's procedure to get the time-dependent invariants.  The
Lagrangian is obtained by making a Legendre transformation to the
Hamiltonian (5) and using the equation of motion (6b), resulting
$$
\eqalign{ L = & {1\over 2} \left\{ A^{-1}_{ \ \ \ \alpha\beta} \dot q_\alpha
\dot q_\beta - (D - CA^{-1} B)_{\alpha\beta} q_\alpha q_\beta +
A^{-1}_{\alpha \beta} F_\alpha F_\beta \right\} \cr
& - (A^{-1} B)_{\alpha\beta} \dot q_\alpha q_\beta -
A^{-1}_{\alpha\beta} F_\alpha \dot q_\beta - G_\alpha q_\alpha +
(A^{-1} B)_{\alpha\beta} F_\alpha q_\beta ~~ . \cr} \eqno(8)
$$
To get the constants of the motion for this  system, let us propose
infinitesimal variations of coordinates given by
$$
\delta q_\alpha = h_\alpha (t) ~~ , \eqno(9)
$$
where ${\bf h} (t)$ is an arbitrary $n-$dimensional vector depending
only on time.  The corresponding variation induced in the Lagrangian
(8) is
$$
\eqalign{ \delta L = & A^{-1}_{\alpha\beta} \dot h_\alpha (\dot
q_\beta - B_{\beta\gamma } q_\gamma - F_\beta) -
(CA^{-1})_{\alpha\beta} h_\alpha \dot q_\beta \cr
& + (CA^{-1} B - D)_{\alpha\beta} h_\alpha q_\beta -
(CA^{-1})_{\alpha\beta} h_\alpha F_\beta - h_\alpha G_\alpha ~~ . \cr
} \eqno(10)
$$
This expression can be rewritten as
$$
\eqalign{ \delta L = & {d\over dt} \left\{ A^{-1}_{\alpha\beta} \dot
h_\alpha q_\beta - (CA^{-1})_{\alpha\beta} h_\alpha q_\beta - \int^t
dt \left(A^{-1}_{\alpha\beta} \dot h_\alpha F_\beta +
(CA^{-1})_{\alpha\beta} h_\alpha F_\beta + h_\alpha G_\alpha\right
)\right  \} \cr
& - \left[\left( A^{-1}_{\alpha\beta} \dot h_\alpha\right)^. + (A^{-1}
B)_{\alpha\beta } \dot h_\alpha - \left( (CA^{-1} )_{\alpha\beta}
h_\alpha \right)^. - (CA^{-1} B-D)_{\alpha\beta} h_\alpha \right]
q_\beta ~~ . \cr} \eqno(11)
$$
Asking that the variations of coordinates (9) yield a Noether's
symmetry transformation, one has that (11) must be equal to a total time
derivative of a function $\Omega_L(q_\alpha,t)$.  This is achieved if the
variation ${\bf h} (t)$ satisfies the differential equation
$$
\eqalign{ & \left( \dot h_\alpha A^{-1}_{\alpha\beta} \right)^ . +
\dot h_\alpha  (A^{-1} B)_{\alpha\beta} \cr
& \quad - \left( h_\alpha (CA^{-1})_{\alpha\beta}\right)^. - h_\alpha
(CA^{-1} B-D ) _{\alpha\beta} = 0 \  , \cr} \eqno(12)
$$
and the function $\Omega_L(q_\alpha,t)$ is defined by the curly brackets of
expression (11). Developing (12), one sees that it has the same form that the
homogeneous part of the differential equation of motion for $ q_{\alpha} $,
given in (7b).  For these symmetry transformations according to the Noether's
theorem the associated conserved charges are given by $J_L = ({\partial {L} /
\partial \dot{q}_\alpha}) \delta q_\alpha - \Omega_L$ , that is
$$
\eqalignno{ J_L = & \quad p_{\alpha} h_\alpha
	- A^{-1}_{\alpha\beta} \dot h_\alpha q_\beta +
	(CA^{-1})_{\alpha\beta} h_\alpha q_\beta \cr
	& + \int^t dt \left( A^{-1}_{\alpha\beta} \dot
	h_\alpha F_\beta - (CA^{-1})_{\alpha\beta}
	h_\alpha F_\beta + h_\alpha G_\alpha \right) \ , &(13) \cr }
$$
with $p_\alpha = \left( A^{-1}_{\alpha\beta} \dot q_\beta - (A^{-1}
B)_{\alpha\beta} q_\beta - A^{-1}_{\alpha\beta} F_\beta \right)$. The
different conserved charges are obtained by the initial conditions
estabished for the system (12).

\vskip 1.0pc

\noindent
{\bf Hamiltonian Formulation}

\vskip 1.0pc

There is also the possibility of using the Noether's theorem in the Hamiltonian
formulation. In this case we consider the action
	$$ S = \int_{t_0}^{t_1}{dt \left\{ \Sigma_{+ \, a b} Q_{b}
		\dot{Q}_{a} - H ( Q_{a} , \, t ) \right\}} \ , \eqno(14) $$

where we have introduced the $(2n \times 2n)$-matrix in block form:
	$$ \Sigma_{+} = \left( \matrix{ 0 & \bf{I} \cr  0 & 0 \cr} \right)  \ ,
		$$
with ${\bf I}$ being the identity matrix in n dimensions.

Taking an arbitrary variation of (14) with respect to coordinates and momenta
defined by (2) we get the expression
	$$ \delta S = \int_{t_0}^{t_1}{ dt \left\{  \delta Q_{a}  \left(
		\Sigma_{+
		\, a b}
		\dot{Q}_{b}  - {\partial \hfill H \over \partial Q_{b} }
		\right) +  \delta
		\dot{Q}_{a}  \Sigma_{- \, a b}  Q_{b} \right\}}
		\  , \eqno(15)  $$
where the transpose of the matrix $ \Sigma_{+}  $ was used,
	$$ \Sigma_{-} \equiv \left( \matrix{ 0 & 0 \cr  \bf{I} & 0 \cr} \right)
		\ . $$
If  the transformation $ \delta Q_{\alpha} $ is a noetherian symmetry , the
curly bracket in (15) must be written as a total time derivative of a function
$ \Omega_H ( Q_{a} , \, t ) $, {\it i.e.},
	$$ \delta Q_{a} \left(  \Sigma_{+ \, a b}  \dot{Q}_{b} -
		{\partial \hfill H
		\over \partial Q_{b} }  \right)  + \delta
		\dot{Q}_{a}  \Sigma_{- \, a b}
		Q_{b}  = {d \over dt}{{\Omega_H}} ( Q_{a} ,  \,   t ) \ .
		\eqno(16) $$
Following now the same procedure as in the Lagrangian case the  associated
conserved quantity to this variation is given by
	$$ J_H = \delta Q_{a} \Sigma_{- \, a b}  Q_{b} -
		\Omega_H  \ . \eqno(17) $$
This can be checked also by taking the total time derivative of (17) and the
Hamilton equations of motion together with the condition (16).

For the multidimensional quadratic parametric Hamiltonian (1), the linear
time-dependent constants of motion can be obtained by means of the variations
	$$ \delta{Q}_{a} = \left( \delta p_{\alpha} , \, \delta q_{\alpha}
		\right) =
		\left( g_{\alpha} (t) \, ,  h_{\alpha} (t) \right) \ . \eqno(18)
		$$
Substituting these variations into the left hand side of (16), we find a
symmetry transformation if the following relations are satisfied:
	$$ \eqalignno{ \dot{g}_{\alpha} &= - C_{\alpha \beta} g_{\beta} -
		D_{\alpha
		\beta} h_{\beta}  \ , & (19a) \cr
		\dot{h}_{\alpha} &= A_{\alpha \beta} g_{\beta} +
		B_{\alpha \beta} h_{\beta}
		 \ , & (19b) \cr
		\Omega_H &= g_{\alpha} q_{\alpha} - \int \, dt \left( F_{\alpha}
		g_{\alpha}
		+ G_{\alpha} h_{\alpha} \right) \ .  & (19c) \cr} $$
It is important to remark that the first two equations have the same form that
the homogeneous part of the Hamilton equations of motion given in (6).
Therefore it is straightforward to demonstrate that the variations $ h_{\alpha}
$ satisfy the same differential equations (12) as the corresponding variations
of the Lagrangian formulation.   For this reason, the constant of motion (17)
coincides with its Lagrangian counterpart (13), {\it i.e.}, $J_L = J_H$.

The expression (13) gives rise to $ 2 n $ invariants correponding to the $ 2 n
$ independent solutions for the system of equations (12).  These $ 2 n $
constants of the motion can be rewritten in matrix form
$$
\pmatrix{ {\bf p}_0 (t) \cr {\bf q}_0 (t) \cr} = \gros\Lambda (t)
\pmatrix { {\bf p} \cr {\bf q} \cr} + \gros\Delta (t) ~~ , \eqno(20)
$$
where the transformation matrix $\gros\Lambda$ takes the form
$$
\gros\Lambda (t) = \pmatrix{ {\bf h}^{(1)} & \left( {\bf h}^{(1)} C -
\dot {\bf h} ^{(1)} \right) A^{-1} \hfill \cr
{\bf h}^{(2)} & \left( {\bf h}^{(2)} C - \dot {\bf h}^{(2)}\right)
A^{-1} \hfill \cr
\vdots & \vdots \cr
{\bf h}^{(2n)} & \left ( {\bf h}^{(2n)} C - \dot {\bf
h}^{(2n)}\right) A^{-1} \hfill \cr} \eqno(21)
$$
The constants of motion defined in (20) satisfy at $t = 0$ the conditions
$\vec{p}_0(0) = \vec{p}$ and $\vec{q}_0(0) = \vec{q }$. These requeriments
imply different initial conditions for the system of differential equations
(12), and guarantee the existence of $2n$ independent solutions, which are
denoted by the superscript on the function ${\bf h}(t)$. Besides these vector
solutions are written horizontally.
The time-dependent column vector $\gros\Delta$ is
given by
$$
\Delta_k (t) = \int^t_0 dt \left( A^{-1}_{\alpha\beta} \dot h
^{(k)}_\alpha F_\beta + (CA^{-1})_{\alpha\beta} h^{(k)}_\alpha
F_\beta + h^{(k)}_\alpha
 G_\alpha \right)  \,  .  \eqno(22)
$$
It is important to remark that the expressions in Eq. (20) are the linear
invariants
proposed in Ref. [8,12], where it was shown that $ {\bf \Lambda} $ is a $2n $
dimensional symplectic matrix, as it must be because the constants of motion
${\bf p}_0 (t)$ and ${\bf q}_0 (t)$ satisfy the canonical commutation
relations. However now the elements of the matrix
$\gros\Lambda$ are given in terms of the $2n$ independent solutions
of the homogeneous classical equations of motion (12), and their
derivatives.

To have that the invariants coincide with the position and momentum
operators at $t = 0$, the appropriate initial conditions for the
solutions of Eq. (12) are
$$
h^{(i)}_j (0) = \cases{ \delta^i \hfill ~~ , & ~~ $1 \leq i, ~  j \leq n$
\hfill
\cr
0 \hfill ~~ , & ~~ $n + 1 \leq i, ~ j \leq 2n$ \cr} \eqno(23a)
$$
and for their derivatives
$$
\dot h^{(i)}_j (0) = \cases{ C_{ij} (0) \hfill ~~ , & ~~ $1 \leq i, ~
j \leq n$ \hfill \cr
-A_{ij} (0) \hfill ~~ , & ~~ $n +1 \leq i, ~ j \leq 2n $ \hfill \cr}
\eqno(23b)
$$

\

\noindent
So we have proved that for multidimensional stationary or non-stationary
classical or quantum systems the linear time-dependent invariants follow from
the Lagrangian or Hamiltonian formulations of the Noether's theorem procedure
by considering a variation
of coordinates or momenta that follows  the classical trajectory.

An additional remark is that the quadratic invariants for the stationary
classical small oscillation system found from non-noetherian transformations
[5] can be obtained from the linear invariants presented here.

The suggested method can be extended to classical or quantum field theory by
considering infinite degrees of freedom and as consequence  finding an infinite
number of time-dependent invariants linear in fields and their corresponding
conjugated momenta.
\

\

\noindent
One of the authors (V.I.M.) would like to thank Instituto de Ciencias
Nucleares, UNAM for its hospitality.

\

\noindent{\bf References}

\

\item{ [1] } T. Takabayashi, {\it Nuovo Cimento} {\bf 33} (1964)
668;\hfill\break
N. Isgur, and G. Karl, {\it Phys. Rev. D \bf 18} (1978) 4187.

\

\item{ [2] } Y. S. Kim, and M. E. Noz, {\it Theory and Applications
of the Poincar\`e Group.} D. Reidel Publishing Company (Dordrecht,
Holland, 1986).

\

\item{ [3] } S. W. Hawking, {\it Comm. Math. Phys. \bf 43} (1975)
199.

\

\item{ [4] }  M. Moshinsky, Canonical Transformations and Quantum
Mechanics. Escuela Latinoamericana de F\ii sica. (Instituto de F\ii
sica, UNAM, M\'exico, 1974); \hfill\break
M. Moshinsky, and P. Winternitz, {\it J. Math. Phys. \bf 21} (1980)
1667.

\

\item{ [5]}  S. Hojman, {\it J. Math. Phys. \bf 34} (1993)
2968.

\
\item{ [6]} V. P. Ermakov, {\it Univ. Izv. Kiev. \bf20 } No. 9 (1880) 1.

\

\item{ [7]}  J. Cari\~nena, J. Fern\'andez-Nu\~nez and E. Mart\ii nez, {\it
Lett. Math. Phys. \bf 23} (1991) 51;
 J. Cari\~nena, E. Mart\ii nez and J. Fern\'andez-Nu\~nez, {\it Reps. Math.
Phys. \bf 31} (1992) 189.

\

\item{ [8] }  V. V. Dodonov and V. I. Man'ko, in: {\it Proc. P. N.
Lebedev Physical Institute}, Vol. 183. Evolution of Multidimensional
Systems. Magnetic Properties of Ideal Gases of Charged Particles, Ed.
M. A. Markov, Nova Science (Commack, N. Y., 1989).

\

\item{ [9] } L. D. Landau and E. M. Lifshitz, {\it Quantum Mechanics,
Nonrelativistic Theory}, Third edition, (Pergamon, Oxford, 1977).

\

\item{ [10] }  H. R. Lewis, {\it Phys. Rev. Lett. \bf 18} (1967) 510,
636; \hfill\break
H. R. Lewis and W. Riesenfeld, {\it J. Math. Phys. \bf 10} (1969)
1458.

\

\item{ [11] } I. A. Malkin, V. I. Man'ko and D. A. Trifonov, {\it
Phys. Lett. A \bf 30} (1969) 414; \hfill\break
I. A. Malkin and V. I. Man'ko, {\it Phys. Lett. A \bf 32}  (1970)
243.

\
\item{ [12] }  I. A. Malkin, V. I. Man'ko and D. A. Trifonov. {\it J.
Math. Phys. \bf 14} (1973) 576.

\

\item{ [13] }  G. Profilo and G Soliani , {\it Phys. Rev. A \bf 44} (1991)
2057;  {\it Ann. Phys.}  (N. Y.)  {\bf 229} (1994) 160.

\

\item{ [14] }  O. Casta\~nos, R. L\'opez-Pe\~na and A. Frank, Noether's theorem
and dynamical symmetries for stationary and non-stationary quantum systems. AIP
Conference Proceedings 266: Group Theory in Physics. Proceedings of the
International Symposium held in Honor of Prof. Marcos Moshinsky,  (Cocoyoc,
Morelos, M\'exico, 1991) p. 109.

\

\item{ [15] }  O. Casta\~nos, R. L\'opez-Pe\~na and V. Man'ko, {\it J. Phys. A
\bf 27} (1994) 1751.

\end